\documentclass[aps,showpacs,showkeys,preprintnumbers,amsmath,amssymb]{revtex4}
\usepackage{mathrsfs, color}
\usepackage{stmaryrd}
\usepackage{cases}
\usepackage{amssymb}
\usepackage{amsmath, amsthm}
\usepackage{amsfonts}
\usepackage{graphicx}
\usepackage{amsmath,amstext,amsbsy,amssymb}
\usepackage{epstopdf}

\newcommand{\ra}{\rangle}
\newcommand{\lan}{\langle}

\def\Tr{\mathrm{Tr}}

\newtheorem{thm}{Theorem}
\newtheorem{lem}{Lemma}
\newtheorem{cor}[thm]{Corollary}
\newtheorem{proposition}{Proposition}

\begin{document}


\title[Enhanced quantum channel uncertainty relations by skew information]{Enhanced quantum channel uncertainty relations by skew information}

\title[Enhanced quantum channel uncertainty relations by skew information]
{Enhanced quantum channel uncertainty relations by skew information}

\author{Xiaoli Hu}
\affiliation{School of Artificial Intelligence, Jianghan University, Wuhan 430056, China}
\email{xiaolihumath@163.com }

\author{Naihong Hu}
\affiliation{School of Mathematical Sciences, Shanghai Key Laboratory of PMMP, East China Normal Unversity, Shanghai 200241, China}
\email{nhhu@math.ecnu.edu.cn}

\author{Bing Yu}
\affiliation{School of Mathematics and Systems Science, Guangdong Polytechnic Normal University,  Guangzhou 510665, China}
\email{mathyu590@163.com}

\author{Naihuan Jing}
\affiliation{Department of Mathematics, North Carolina State University, Raleigh, NC 27695, USA}
\email{jing@ncsu.edu}

\date{\today}

\begin{abstract}By revisiting the mathematical foundation of the uncertainty relation, skew information-based uncertainty sequences are developed for any two quantum channels. A reinforced version of the Cauchy-Schwarz inequality is adopted to improve the uncertainty relation, and a sampling technique of observables' coordinates is used to offset randomness in the inequality. It is shown that the lower bounds of the uncertainty relations are tighter than some previous studies.
\end{abstract}



\maketitle
\section{Introduction}\label{sec1}
The uncertainty relation, which is at the core of quantum theory, has a wide range of applications in quantum information and quantum computation such as quantum cryptography, quantum entanglement, the quantum speed limit, signal processing etc. \cite{1,RB,3,4,5,6,7}.
A crucial role was performed \cite{15} by the well-known Heisenberg uncertainty principle, which offered profound insights into the nature of the quantum world and set it apart from the classical one. The classical Heisenberg uncertainty relation states that it is impossible to precisely measure both momentum $p$ and position $x$ at the same time. Weyl and Kennard \cite{16,17} have both given  the present formulation of the uncertainty principle. Robertson demonstrated \cite{18} that the product of the standard deviations  has a generic lower bound for any pair of non-commuting bounded observables $A$ and $B$:
\begin{equation}\label{eq:Rober 2}\Delta A\Delta B\geq |\lan [A,B]\ra|/2,
\end{equation}
where $\Delta$ is the  standard deviation and $\lan [A,B]\ra$ is the expectation value of the commutator $[A, B]=AB-BA$ with respect to a quantum state.
The uncertainty relations are usually cast in terms of mutual information \cite{GH}, Shannon entropy \cite{DD,MHU,WY}, R\'enyi entropy \cite{MHU}, conditional entropy \cite{RB, GG}, and other metrics as well as variance. In most cases, the quantum state and the observables are both important in determining the uncertainty relations.

As the broadest form of measurement \cite{BG,NC}, quantum channels are crucial for processing quantum information in quantum theory \cite{XJ,MK,TRA}.
The uncertainty relation for quantum channels places a fundamental limit on how much information may pass via a quantum channel with a given level of noise, where the uncertainty of the product of channel capacity and channel noise level reveals
the quantum channel's fundamental properties. This is analogous to Heisenberg's uncertainty principle in quantum physics which sets a basic limit on the accuracy
in measuring a pair of physical quantities. The design and analysis of quantum communication protocols, such as quantum key distribution and quantum teleportation, heavily rely on the uncertainty relation for quantum channels. In addition, skew information was also used in the uncertainty relation for general quantum channels. For instance, the parallelogram rule was employed to construct the uncertainty relation for two quantum channels \cite{FSL} and
the norm inequality of Hilbert space was adopted to expand the uncertainty relation for multiple quantum channels \cite{LS18,ZWF}.

The concept of the skew information was originally established by Wigner and Yanase from an information-theoretic perspective. It is a metric for the noncommutativity between the observable $A$ and the quantum state $\rho$:
\begin{equation}
	I(\rho, A)=-\frac{1}{2}\Tr [\sqrt{\rho}, A]^2.
\end{equation}

For pure states, the skew information and variance coincide, but the former more suitable in representing the quantum character \cite{LS05}. To be more precise, the skew information measures the non-commutativity between two quantum observables, and captruring the quantum correlations and non-classical properties of a system. Moreover, the skew information can be seen as a true indicator of quantum coherence for a quantum state \cite{YC}, thus it has been extensively studied in
 quantum information theory. In \cite{ZG, LG}, the authors gave out tighy sum-form uncertainty relation by skew information for observables and channels. For any two observables $A$ and $B$, the uncertainty relation based on the skew information \cite{LS03} says that
 \begin{equation}
 I(\rho, A)I(\rho, B)\geq \frac{1}{4}\bigg|\Tr \rho [A,B]\bigg|^2.
 \end{equation}

In this paper, we aim to provide improved uncertainty relations in skew-information for two arbitrary quantum channels by utilizing a new strengthened form of the
Cauchy-Schwarz inequality. Our methodology in tightening the inequality relies on a recently known technique \cite{HJ23} and can be presented in an easy way. We also
 use symmetry to count for the randomness of the measured coordinates. It turns out that the new lower bounds outperform some of the recent studies (eg. \cite{ZZ}).

The structure of this article is as follows.  In section II, we strengthen the Cauchy-Schwarz inequality and provide solid uncertainty relations
for the product and sum form of two quantum channels using a ``fine-grained'' sequence of inequalities.  The symmetry of permutations is applied to strengthen and offset
the randomness in the uncertainty related sequence in section III. We also show that the lower bounds are independent from the Kraus representations of quantum channels. Summary and conclusions are given in Section VI.

\section{THE UNCERTAINTY RELATIONS FOR QUANTUM CHANNELS BY SKEW INFORMATION}\label{sec2}

A general quantum state of a $d$-dimensional quantum system $H$ is presented by a density matrix $\rho$, which is a positive semidefinite operator
 on the Hilbert space $H\cong \mathbb{C}^d$ with trace one. The inner product for operators $X$ and $Y$ acting on $H$ is defined as $(X, Y) = \Tr (X^{\dagger} Y)$.
  A quantum channel $\mathcal{N}$, a completely positive (CP) and trace preserving (TP) linear map from operators on $GL(H_1)$ to operators on $GL(H_2)$, best captures the evolution of quantum systems. Through the work of Kraus (1971) and Choi (1975), it has been known that a generic channel admits a Kraus representation so that its action on a density matrix $\rho$ on $H$ takes the form $\mathcal{N}(\rho)=\sum_iK_i\rho K_i^{\dagger}$, where $\{K_i\}$ is a set of operators such that $\sum_i K_iK_i^{\dagger}=I$. The largest number of Kraus operators required for a Kraus representation of $\mathcal{N}$ is equal to the dimension of $Hom(H_1, H_2)$. For example, the maximum number of Kraus operators for the case $ H_1\cong H_2\cong \mathbb{C}^d$ is $d^2$. Therefore, we can write $\mathcal{N}(\rho)=\sum_{i=1}^n  K_i\rho K_i^{\dagger}$ with $\sum_{i=1}^n K_iK_i^{\dagger}=I$, where $n$ is a natural number that does not exceed $d^2$.

The skew information of an arbitary quantum channel $\mathcal{N}$ with respect to a quantum state $\rho$ can be written as the sum of the skew informations of the Kraus operators $K_i$. That is 
\begin{equation}
	I(\rho,\mathcal{N})=\sum_{i=1}^n I(\rho,K_i),
\end{equation}
where $I(\rho ,K_i)=\frac{1}{2}\Tr\big([\sqrt{\rho}, K_i]^{\dagger }[\sqrt{\rho}, K_i]\big)$
  is the skew information of the $i$-th Kraus operator with respect to the quantum state $\rho$ \cite{SY}. Here $I(\rho, \mathcal{N})$ satisfies a number of desired features and is a legitimate measure of both coherence and quantum uncertainty of $\rho$ with respect to quantum channel $\mathcal{N}$.

   \subsection{Main results}

\begin{lem} Let $\rho$ be a quantum state in  a $d$-dimentional Hilbert space $H$. For two quantum channels $\mathcal{N}_1$ and $\mathcal{N}_2$  with  Kraus decompositions $\mathcal{N}_1=\sum_{i=1}^{n_1}E_i\rho E_i^{\dagger}$ and $\mathcal{N}_2=\sum_{j=1}^{n_2}F_j\rho F_j^{\dagger}$ respectively, the following uncertainty relation holds:
  \begin{equation}
 	I(\rho, \mathcal{N}_1)I(\rho,\mathcal{N}_2)\geq  \frac{1}{4}\sum_{i=1}^{n_1}\sum_{j=1}^{n_2} \big([\sqrt{\rho},E_i],[\sqrt{\rho},F_j]\big)^2,
 \end{equation}
where  $\big([\sqrt{\rho},E_i],[\sqrt{\rho},F_j]\big)=\Tr\bigg([\sqrt{\rho},E_i]^{\dagger}[\sqrt{\rho},F_j]\bigg) $ represents the inner product for operators that are acting on $H$.
\end{lem}
\proofname:
  Setting $\rho^{E_i}:=[\sqrt{\rho},E_i]$, $\rho^{F_j}:=[\sqrt{\rho},F_j]$ and $|\rho^{E_i}_k\rangle:=\rho^{E_i}|k\rangle$, $|\rho^{F_j}_l\rangle:=\rho^{F_j}|l\rangle$, we have
\begin{equation}\label{eq:1}
	\begin{split}
		I(\rho,E_i)I(\rho,F_j)=&\frac{1}{4}\Tr\bigg([\sqrt{\rho}, E_i]^{\dagger }[\sqrt{\rho}, E_i]\bigg)\cdot \Tr\bigg([\sqrt{\rho}, F_j]^{\dagger }[\sqrt{\rho}, F_j]\bigg)\\
	=&	\frac{1}{4}\Tr\bigg( (\rho^{E_i})^{\dagger}\rho^{E_i}\bigg)\cdot \Tr\bigg( (\rho^{F_j})^{\dagger}\rho^{F_j}\bigg) \\
	=&\frac{1}{4}\bigg(\sum_{k=1}^d\langle k| (\rho^{E_i})^{\dagger}\rho^{E_i}|k\rangle\bigg) \cdot \bigg(\sum_{l=1}^d\langle l| (\rho^{F_j})^{\dagger}\rho^{F_j}|l\rangle\bigg)\\
	=&\frac{1}{4} \bigg( \sum_{k=1}^d \langle \rho^{E_i}_k|\rho^{E_i}_k\rangle \bigg) \bigg(\sum_{l=1}^d \langle \rho^{F_j}_l |\rho^{F_j}_l\rangle \bigg).
			\end{split}
\end{equation}
The third equation follows from the fact that matrices $(\rho^{E_i})^{\dagger}\rho^{E_i}$ and $(\rho^{F_j})^{\dagger}\rho^{F_j}$ are hermitian, ensuring that their eigenvalues are real.
Viewing $\vec{e}^{~i}=(|\rho^{E_i}_1\ra, |\rho^{E_i}_2\ra,\cdots,|\rho^{E_i}_d\ra)$ and $\vec{f}^{~j}=(|\rho^{F_j}_1\ra, |\rho^{F_j}_2\ra,\cdots,|\rho^{F_j}_d\ra)$ as two vectors in $(\mathbb{C}^{d})^d$ and  using the Cauchy-Schwarz inequality, we get 
\begin{equation}\label{e:7}
	\begin{split}
I(\rho,E_i)I(\rho,F_j)=&\frac{1}{4}\lan \vec{e}^{~i},  \vec{e}^{~i}\ra\lan \vec{f}^{~j},  \vec{f}^{~j}\ra\\
\geq &\frac{1}{4}|\lan \vec{e}^{~i},\vec{f}^{~j}\ra|^2
	= \frac{1}{4}\left|\sum_{k=1}^d \langle\rho^{E_i}_k|\rho^{F_j}_k\rangle\right|^2\\
	=&
	\frac{1}{4}\Tr\bigg([\sqrt{\rho},E_i]^{\dagger}[\sqrt{\rho},F_j]\bigg)\\
	=&\frac{1}{4}\big([\sqrt{\rho},E_i], [\sqrt{\rho},F_j]\big)^2,
	\end{split}
	\end{equation}
where $\lan , \ra$ is the inner product on $(\mathbb{C}^{d})^d$. Taking sum of the $i$ and $j$ terms on both sides of \eqref{e:7}, we have that
\begin{equation}
		I(\rho,\mathcal{N}_1)I(\rho,\mathcal{N}_2)=\sum_{i=1}^{n_1}\sum_{j=1}^{n_2}I(\rho,E_i)I(\rho,F_j)\geq \frac{1}{4}\sum_{i=1}^{n_1}\sum_{j=1}^{n_2}  \big([\sqrt{\rho},E_i], [\sqrt{\rho},F_j]\big)^2.
\end{equation}
This completes the proof. 	$\hfill\qed$
\bigskip

For $m\in \{1,2,\cdots, d\}$, we denote by $\vec{e}^{~i}_m=(|\rho^{E_i}_1\ra,\cdots,|\rho^{E_i}_m\ra,0,\cdots,0)^{\top}$ the partial vector of $\vec{e}^{~i}$,  and $\vec{e}^{~i}_{m,c}=(0,\cdots,0,|\rho^{E_i}_{m+1}\ra\cdots,|\rho^{E_i}_d\ra)^{\top}$ the complementary of the vector $\vec{e}^{~i}_m$, where $\top$ is the matrix transpose. Similarly $\vec{f}^{~j}_m$ and $\vec{f}^{~j}_{m,c}$ are the partial vector and the complementary for $\vec{f}^{~j}_m$.
To enhance the uncertainty relation in Lemma 1, we tighten the Cauchy-Schwarz inequality using the following technique (cf. \cite{HJ23}):
 \begin{equation}\label{eq:Iij}
 	\begin{split}
 		I(\rho,E_i)I(\rho,F_j)=&\frac{1}{4}|\vec{e}^{~i}|^2|\vec{f}^{~j}|^2=\frac{1}{4}(|\vec{e}^{~i}_m|^2+|\vec{e}^{~i}_{m,c}|^2)(|\vec{f}^{~j}_m|^2+|\vec{f}^{~j}_{m,c}|^2)\\
 		=&\frac{1}{4} \bigg (|\vec{e}^{~i}_m|^2|\vec{f}^{~j}_m|^2+|\vec{e}^{~i}_m|^2|\vec{f}^{~j}_{m,c}|^2+ |\vec{e}^{~i}_{m,c}|^2(|\vec{f}^{~j}_m|^2+|\vec{f}^{~j}_{m,c}|^2)\bigg)\\
 		\geq &\frac{1}{4} \bigg (|\lan\vec{e}^{~i}_m, \vec{f}^{~j}_m\ra |^2+|\vec{e}^{~i}_m|^2|\vec{f}^{~j}_{m,c}|^2+ |\vec{e}^{~i}_{m,c}|^2(|\vec{f}^{~j}_m|^2+|\vec{f}^{~j}_{m,c}|^2)\bigg).
 	\end{split}
 	 \end{equation}
Let us denote
\begin{equation}\label{Imij}
 I_m^{(i,j)}:=\frac{1}{4} \bigg (|\lan\vec{e}^{~i}_m, \vec{f}^{~j}_m\ra |^2+|\vec{e}^{~i}_m|^2|\vec{f}^{~j}_{m,c}|^2+ |\vec{e}^{~i}_{m,c}|^2(|\vec{f}^{~j}_m|^2+|\vec{f}^{~j}_{m,c}|^2)\bigg),
\end{equation}
where $m=1,2,\cdots, d$.

In particular, we have $I_1^{(i,j)}=I(\rho,E_i)I(\rho, F_j)$ and
$I_{d}^{(i,j)}=\frac{1}{4} |\lan \vec{e}^{~i},\vec{f}^{~j}\ra|^2=\frac{1}{4} \big([\sqrt{\rho},E_i], [\sqrt{\rho},F_j]\big)^2$.

\begin{thm} Let $\rho $ be a quantum state on a $d$-dimentional Hilbert space $H$, and let $\mathcal{N}_1$ and $\mathcal{N}_2$ be two quantum channels with the Kraus decompositions $\mathcal{N}_1(\rho)=\sum_{i=1}^{n_1}E_i\rho E_i^{\dagger}$ and $\mathcal{N}_2(\rho)=\sum_{j=1}^{n_2}F_j\rho F_j^{\dagger}$ respectively. Then we have the following uncertainty sequance:
	\begin{equation} \label{eq:th1}
	I(\rho,\mathcal{N}_1)I(\rho,\mathcal{N}_2)= I_1\geq I_2\geq \cdots \geq I_{d}=\frac{1}{4}\sum_{i=1}^{n_1}\sum_{j=1}^{n_2} \big([\sqrt{\rho},E_i], [\sqrt{\rho},F_j]\big)^2,
	\end{equation}
where $I_m=\sum_{i=1}^{n_1}\sum_{j=1}^{n_2}I_m^{(i,j)}$  for $1\leq m\leq d$.
\end{thm}
\proofname: By (\ref{eq:Iij}) and (\ref{Imij}),  we have
$I(\rho,E_i)I(\rho,F_j)\geq I_m^{(i,j)}$. It is readily seen that
	\begin{equation}\label{eq:th21}
		I^{(i,j)}_{m-1}-I^{(i,j)}_m=\bigg(\sum_{k=1}^{m-1}\langle \rho_k^{E_i}|\rho_m^{F_j}\rangle +\langle \rho_m^{E_i}|\rho_k^{F_j}\rangle \bigg)^2\geq 0,
	\end{equation}
therefore
	\begin{equation} \label{eq:th22}
		I(\rho,E_i)I(\rho,F_j)=I_1^{(i,j)}\geq I_2^{(i,j)}\geq \cdots\geq I_{d}^{(i,j)}=\frac{1}{4}\big([\sqrt{\rho},E_i], [\sqrt{\rho},F_j]\big)^2.
	\end{equation}
Taking sum over $i$ and $j$, we get (\ref{eq:th1}). $\hfill\qed$
\bigskip

The following is an immediate consequence of Theorem 1:
\begin{cor}
	Let $\rho$ be a quantum state on a $d$-dimentional Hilbert space $H$, and let $\mathcal{N}_1$ and $\mathcal{N}_2$ be two quantum channels with Kraus decompositions $\mathcal{N}_1(\rho)=\sum_{i=1}^{n_1}E_i\rho E_i^{\dagger}$ and $\mathcal{N}_2(\rho)=\sum_{j=1}^{n_2}F_j\rho F_j^{\dagger}$ respectively.
	Then the following holds:
	\begin{equation}
		\begin{split}
		I(\rho, \mathcal{N}_1)+I(\rho,\mathcal{N}_2)&\geq 2\sqrt{I_1}\geq 2\sqrt{I_2} \geq2\sqrt{I_3}\geq\cdots\\
		&\geq 2\sqrt{I_{d}}=2\bigg (\sum_{i=1}^{n_1}\sum_{j=1}^{n_2} \big([\sqrt{\rho},E_i], [\sqrt{\rho},F_j]\big)^2\bigg)^{\frac{1}{2}},	
	\end{split}
	\end{equation}
	where $I_m=\sum_{i=1}^{n_1}\sum_{j=1}^{n_2}I_m^{(i,j)}$ for $1\leq m\leq d$.
\end{cor}

Next we strengthen the Cauchy-Schwarz inequality by introducing another sequence for the uncertainty relation.
For $1\leq q<p\leq d$, set $S_{1,0}^{(i,j)}:=I(\rho,E_i)I(\rho,F_j)$ and (cf. \cite{HJ23})
\begin{equation}\label{eq:S}
	\begin{split}
		S_{p,q}^{(i,j)}&:=S_{p,q-1}^{(i,j)}-\bigg(\langle \rho^{E_i}_p|\rho^{E_i}_p\rangle + \langle \rho^{F_j}_q|\rho^{F_j}_q\rangle)\bigg)+\bigg(\langle \rho^{E_i}_p|\rho^{F_j}_p\rangle +\langle \rho^{E_i}_q| \rho^{F_j}_q\rangle\bigg)^2,\\
			S_{p+1,1}^{(i,j)}&:=S_{p,p-1}^{(i,j)}-
			\bigg(\langle \rho^{E_i}_{p+1}|\rho^{E_i}_{p+1}\rangle + \langle \rho^{F_j}_1|\rho^{F_j}_1\rangle\bigg)+\bigg(\langle \rho^{E_i}_{p+1}|\rho^{F_j}_{p+1}\rangle +\langle \rho^{E_i}_1| \rho^{F_j}_1\rangle\bigg)^2.
		\end{split}
\end{equation}
 Specially,
$S^{(i,j)}_{d,d-1}=\frac{1}{4}\big([\sqrt{\rho},E_i], [\sqrt{\rho},F_j]\big)^2$.
\begin{thm} Let $\rho $ be a quantum state on a $d$-dimentional Hilbert space $H$.
For arbitary two quantum channels $\mathcal{N}_1$ and $\mathcal{N}_2$ with the Kraus decompositions $\mathcal{N}_1(\rho)=\sum_{i=1}^{n_1}E_i\rho E_i^{\dagger}$ and $\mathcal{N}_2(\rho )=\sum_{j=1}^{n_2}F_j\rho F_j^{\dagger}$ respectively. We have the following uncertainty sequence:
\begin{equation}\label{eq:th3}
	\begin{split}
	I(\rho,\mathcal{N}_1)I(\rho,\mathcal{N}_2)&= S_{1,0}\geq S_{2,1}=I_2\geq S_{3,1}\geq  S_{3,2}=I_3\geq S_{4,1}\geq S_{4,2}\\ &\geq S_{4,3}=I_4\geq\cdots
	\geq S_{d,d-1}=\frac{1}{4}\sum_{i=1}^{n_1}\sum_{j=1}^{n_2} \big([\sqrt{\rho},E_i], [\sqrt{\rho},F_j]\big)^2,
\end{split}
\end{equation}
where  $S_{p,q}=\sum_{i=1}^{n_1}\sum_{j=1}^{n_2}S_{p,q}^{(i,j)}$ with  $1\leq q< p\leq d$.
\end{thm}
\proofname: It follows from (\ref{eq:S}) that
\begin{equation}\label{eq:Sij}
	S_{1,0}^{(i,j)}\geq S^{(i,j)}_{2,1}\geq S_{3,1}^{(i,j)}\geq  S_{3,2}^{(i,j)}\geq S_{4,1}^{(i,j)}\geq S_{4,2}^{(i,j)}\geq S_{4,3}^{(i,j)}\geq  \cdots
	\geq S^{(i,j)}_{d,d-1}.
\end{equation}
Then (\ref{eq:th3}) follows from taking sum of (\ref{eq:Sij}) over $i$ and $j$. This completes the proof. $\hfill\qed$
\bigskip
\begin{cor}
	Let $\rho$ be a quantum state on a $d$-dimentional Hilbert space, and let $\mathcal{N}_1$ and $\mathcal{N}_2$ be two quantum channels with Kraus decompositions $\mathcal{N}_1(\rho)=\sum_{i=1}^{n_1}E_i\rho E_i^{\dagger}$ and $\mathcal{N}_2(\rho)=\sum_{j=1}^{n_2}F_j\rho F_j^{\dagger}$ respectively. Then we have that
	\begin{equation} \label{eq:sum}
		\begin{split}
			I(\rho, \mathcal{N}_1)+I(\rho,\mathcal{N}_2)&\geq 2\sqrt{S_{1,0}}\geq 2\sqrt{S_{2,1}}\geq 2\sqrt{S_{3,1}} \geq 2\sqrt{S_{3,2}}\geq\cdots \sqrt{S_{p,q}}\geq\cdots \\
			&\geq 2\sqrt{S_{d,d-1}}=\bigg(\sum_{i=1}^{n_1}\sum_{j=1}^{n_2}\big([\sqrt{\rho},E_i], [\sqrt{\rho},F_j]\big)^2\bigg)^{\frac{1}{2}},
		\end{split}
	\end{equation}
	where $S_{p,q}=\sum_{i=1}^{n_1}\sum_{j=1}^{n_2}S_{p,q}^{(i,j)}$ with $1\leq q< p\leq d$.
\end{cor}

\subsection{An example}
We consider the following decomposable quantum state $\rho(\theta )$ in the Hilbert space $\mathbb{C}^4$,
\begin{equation}\label{rho}
\rho(\theta)=\frac{1}{4}	
\left(\begin{array}{cccc}
	1 & 2 \theta -1 & 0 & 0 \\
	 2 \theta -1 & 1 & 0 & 0 \\
		0 & 0 & 1 & 2 \theta -1 \\
		0 & 0 &  2 \theta -1 & 1 \\
	\end{array}
	\right), \qquad 0\leq \theta \leq 1,
\end{equation}
Let $\mathcal{N}_1$ and $\mathcal{N}_2$ be the quantum channels with the Kraus operators $\{E_i\}$ and $\{F_i\}$ respectively:
\begin{equation}\label{K}
	\begin{split}
E_1=&\left(
\begin{array}{cccc}
	1 & 0 & 0 & 0 \\
	0 & \sqrt{1-p} & 0 & 0 \\
	0 & 0 & 1 & 0 \\
	0 & 0 & 0 & \sqrt{1-p} \\
\end{array}
\right), \quad E_2=\left(
\begin{array}{cccc}
	0 & 0 & 0 & 0 \\
	0 & \sqrt{p} & 0 & 0 \\
	0 & 0 & 0 & 0 \\
	0 & 0 & 0 & \sqrt{p} \\
\end{array}
\right), \\
\\
F_1=&\left(
\begin{array}{cccc}
	\sqrt{1-q} & 0 & 0 & 0 \\
	0 & 1 & 0 & 0 \\
	0 & 0 &\sqrt{1-q}& 0 \\
	0 & 0 & 0 &1 \\
\end{array}
\right), \quad F_2=\left(
\begin{array}{cccc}
	0 & \sqrt{q} & 0 & 0 \\
	0 & 0 & 0 & 0 \\
	0 & 0 & 0 & \sqrt{q} \\
	0 & 0 & 0 & 0 \\
\end{array}
\right).
\end{split}
\end{equation}
where $0\leq p, q\leq 1$.

When $\theta=\frac{1}{2}$, the quantum state $\rho(\theta)$ is degenerated to an incoherent state $\frac{1}{2}(|0\ra\lan0|+|1\ra\lan1|)\otimes \frac{1}{2}(|0\ra\lan0|+|1\ra\lan1|)$ which is swapped with any operators. The skew information $I(\rho(\theta), E_i)$ is a metric for the noncummutativity between $\rho$ and operators $E_i$. Hence $I(\rho(\theta),\mathcal{N})$ tends to 0 when  $\theta$ approaches to $\frac{1}{2}$. When $\theta=1$ (or $\theta=0$), $\rho(\theta)$ is the tensor product of the incoherent state $\frac{1}{2}(|0\ra\lan0|+|1\ra\lan1|)$ and the maximally coherent state $\frac{1}{2}(|0\ra\lan0|+|0\ra\lan1|+|1\ra\lan0|+|1\ra\lan1|)$ (or $\frac{1}{2}(|0\ra\lan0|-|0\ra\lan1|-|1\ra\lan0|+|1\ra\lan1|)$).  The following calculation will show that $I(\rho(\theta),\mathcal{N})$ will attain the maximal value in these cases.
Direct computation  gives that
\begin{align}\label{eq:MP}
		I(\rho,\mathcal{N}_1)I(\rho,\mathcal{N}_2)&=\frac{1}{4}\left(\sqrt{1-\theta }-\sqrt{\theta }\right)^4 \left(1-\sqrt{1-p}\right) \left(1-\sqrt{1-q}\right),\\ \label{eq:MS}
			I(\rho, \mathcal{N}_1)+I(\rho, \mathcal{N}_2)&=\left(2 \sqrt{-(\theta -1) \theta }-1\right) \left(\sqrt{1-p}+\sqrt{1-q}-2\right),\\ \label{eq:LBZ}
			LBZ &=\frac{1}{8} \left(\sqrt{1-\theta }-\sqrt{\theta }\right)^4 \left(1-\sqrt{1-p}\right) \left(1-\sqrt{1-q}\right)^2,
\end{align}
where $LBZ=\frac{1}{4}\sum_{i,j=1}^2 \big([\sqrt{\rho},E_i], [\sqrt{\rho},F_j]\big)^2$ is the latest lower bound in \cite{ZZ}.

The low bounds $I_2$ and $I_3$ in Theorem 1, and the lower bounds $S_{21}, S_{31}$ and $S_{32}$ in Theorem 2 are given by
\begin{equation}\label{eq:S21}
		I_2=S_{21}=\frac{1}{32} \left(1-2 \sqrt{-(\theta -1) \theta }\right)^2 \left(\sqrt{1-p}-1\right) \left(q+8 \sqrt{1-q}-8\right),\\
\end{equation}
\begin{equation}\label{eq:S31}
	\begin{split}
		S_{31}=&\frac{1}{256}\left(4 \theta ^2-4 \theta +4 \sqrt{-(\theta -1) \theta }-1\right)\cdot\\
		& \left(p \left(q+2 \sqrt{1-q}-2\right)-8 \left(\sqrt{1-p}-1\right) \left(2 q+9 \sqrt{1-q}-9\right)\right),
	\end{split}
\end{equation}
\begin{equation}\label{eq:S32}
	\begin{split}
		I_3=S_{32}=&\frac{3q}{256} \left(\sqrt{1-\theta }-\sqrt{\theta }\right)^4 \left(\sqrt{1-p}-1\right)^2 \\
		&+\frac{1}{16} \left(1-2 \sqrt{(1-\theta ) \theta }\right)^2 \left(\sqrt{1-p}-1\right)^2 \left(\sqrt{1-q}-1\right)^2\\
		&+\frac{p}{16} \left(1-2 \sqrt{(1-\theta ) \theta }\right)^2  \left(\sqrt{1-q}-1\right)^2\\
		&+\frac{q}{16} \left(1-2 \sqrt{(1-\theta ) \theta }\right)^2 \left(\sqrt{1-p}-1\right)^2 \\
		&+\frac{q\sqrt{p}}{256} \left(1-2 \sqrt{(1-\theta ) \theta }\right)^2 \left(4 \sqrt{1-p}+3 \sqrt{p}-4\right) .\\
			\end{split}
\end{equation}

The lower bounds of the product-form and sum-form skew information uncertainty relations are given in figure 1 and figure 2 respectively when $\theta=1$.

\begin{figure}
	\includegraphics[width=12.5cm,height=7cm]{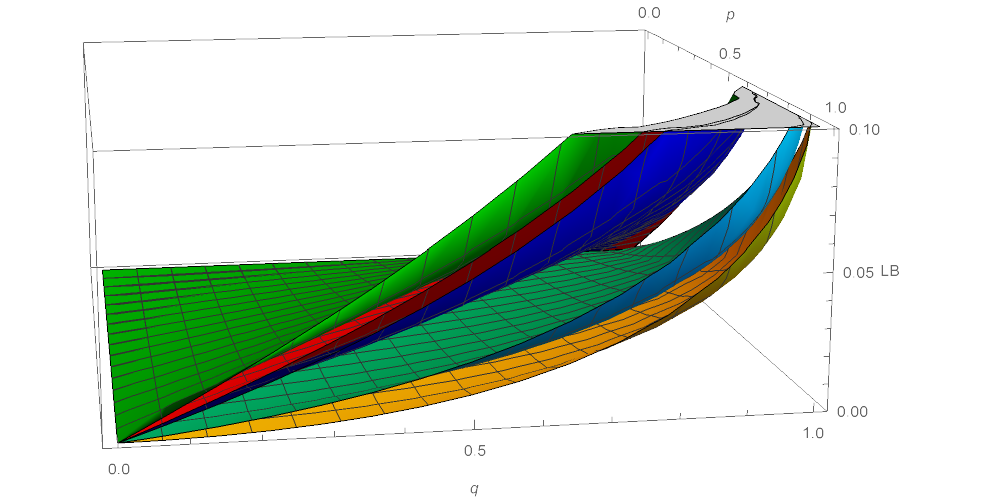}
		\caption{(color online) \textbf{The lower bounds (LB) of $I(\rho, \mathcal{N}_1)I(\rho,\mathcal{N}_2)$ for the state $\rho$ with $\theta= 1$.} Here $p$ and $q$ are the parameters of the Kraus representations for $\mathcal{N}_1$ and $\mathcal{N}_2$ respectively. The
			green, red, blue and cyan  surfaces (the 1st, 2nd, 3rd and 4th surfaces from top to bottom) are the product $I(\rho, \mathcal{N}_1)I(\rho,\mathcal{N}_2)$ in (\ref{eq:MP}),  the lower bound $S_{21}=I_2$ in (\ref{eq:S21}), the lower bound $S_{31}$ in (\ref{eq:S31}), and the lower bound $S_{32}=I_3$ in (\ref{eq:S32}) respectively. The yellow surface (the bottom one) is the lower bound of \cite{ZZ} (i.e. LBZ in (\ref{eq:LBZ})).}
\end{figure}
\begin{figure}
	\includegraphics[width=14cm,height=7cm]{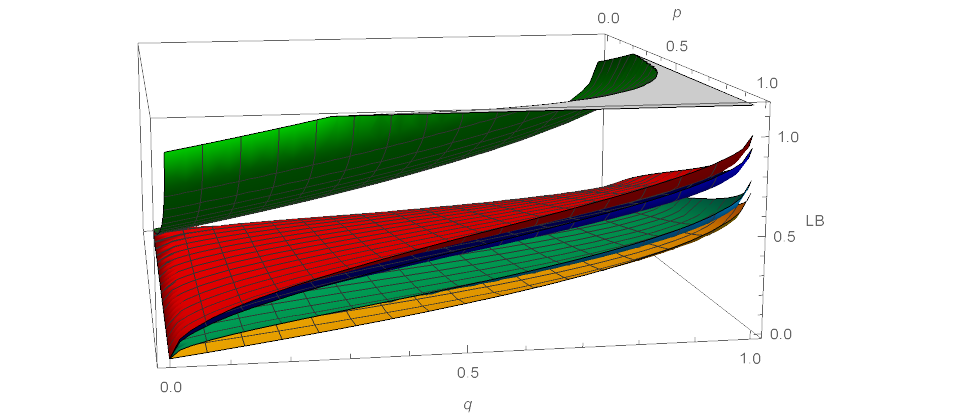}
	\caption{(color online) \textbf{The lower bounds (LB) of the sum $I(\rho, \mathcal{N}_1)+I(\rho,\mathcal{N}_2)$ for the state $\rho$ with $\theta= 1$.} Here $p$ and $q$ are the parameters of the Kraus representations for $\mathcal{N}_1$ and $\mathcal{N}_2$ respectively. The green, red, blue and cyan surfaces (the 1st, 2nd, 3rd and 4th surface from top to bottom)
 represent the product $I(\rho, \mathcal{N}_1)+I(\rho,\mathcal{N}_2)$ in (\ref{eq:MS}), the lower bound $2\sqrt{S_{21}}$, the lower bound $2\sqrt{S_{31}}$, and
the lower bound $2\sqrt{S_{32}}$ respectively. The yellow surface (the bottom surface) is the lower bound $2\sqrt{\hbox{LBZ}}$.}
\end{figure}
From figure 1, we can see that our lower bound $S_{21}=I_2$ (the second surface from top to bottom and in red) is the closest one to the product $I(\rho, \mathcal{N}_1)I(\rho,\mathcal{N}_2)$ (the top surface in green). Our bounds $S_{21}, S_{31}, S_{32}$ (the 2nd, 3rd and 4th surfaces from top to bottom) are better than the latest lower bound LBZ (the bottom surface in yellow) in \cite{ZZ}. Figure 2 also shows that our bounds of the sum form skew information for two quantum channels are better than the lower bound in \cite{ZZ}. However, the gap between them and the sum $I(\rho, \mathcal{N}_1)+I(\rho,\mathcal{N}_2)$ is still large. In the next section, we propose improved uncertainty relations.

\begin{figure}
	\includegraphics[width=7.5cm,height=5cm]{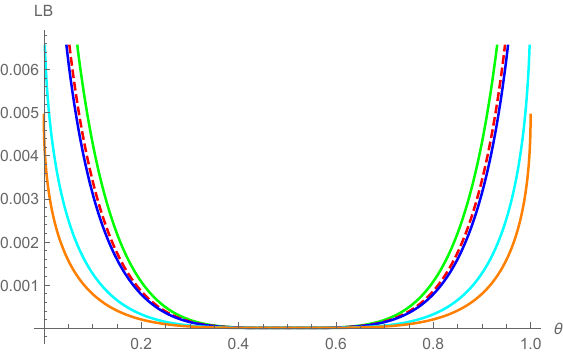}
	\quad
	\includegraphics[width=7.5cm,height=5cm]{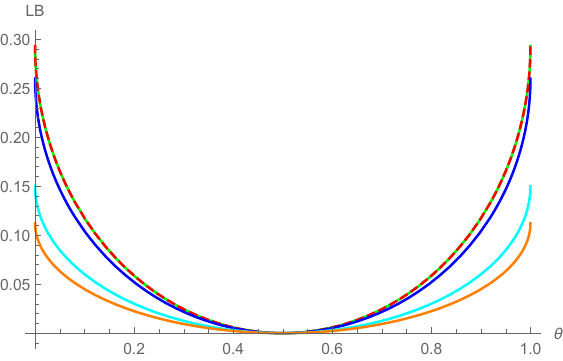}
	\caption{(color online) \textbf{The lower bounds (LB) of the product $I(\rho, \mathcal{N}_1)I(\rho,\mathcal{N}_2)$ (the left figure) and the sum $I(\rho, \mathcal{N}_1)+I(\rho,\mathcal{N}_2)$ (the right figure) for the quantum channels $\mathcal{N}_1$ and $\mathcal{N}_2$ with the Kraus operators in (\ref{K}) with $p=q=0.5$.} Here $\theta$ is the parameter of the quantum state $\rho(\theta)$. The left and right green curves are the product $I(\rho, \mathcal{N}_1)I(\rho,\mathcal{N}_2)$ in (\ref{eq:MP})and the sum form $I(\rho, \mathcal{N}_1)+I(\rho,\mathcal{N}_2)$ in (\ref{eq:MS}),  the left and right dashed red curves are the lower bounds $S_{21}$ and $2\sqrt{S_{21}}$ respectively, the left and right blue curves are the lower bounds $S_{31}$ and $2\sqrt{S_{31}}$ respectively, the left and right cyan curves are the lower bounds $S_{32}$ and $2\sqrt{S_{32}}$, the left and right yellow curves are the lower bounds LBZ and $2\sqrt{\hbox{LBZ}}$ respectively.}
\end{figure}

From figure 3 we can see that both the product $I(\rho, \mathcal{N}_1)I(\rho,\mathcal{N}_2)$ and the sum $I(\rho, \mathcal{N}_1)+I(\rho,\mathcal{N}_2)$ approach to 0 as $\theta$ tends to $\frac{1}{2}$. As we explained above,
the noncommutability of the quantum channels  $\mathcal{N}_1, \mathcal{N}_2$ and the quantum state $\rho(\theta)$ tends to 0 in this case.

\section{Improved uncertainty relations independent of specific Kraus decompositions}\label{sec3}
As Theorem 3 is a better version of Theorem 1, we are focused on improving the uncertainty relationship in Theorem 3.

The lower bounds of Theorem 3 can be strengthened by using the symmetric group $\mathfrak{S}_d$, which naturally acts by permuting the set $\{1, 2,..., d\}$. The induced action of $\mathfrak{S}_d\times \mathfrak{S}_d$ on $S^{(i,j)}_{p,q}$ and $S^{(i,j)}_{p+1,1}$ in (\ref{eq:S}) for any two permutations $\sigma,\tau \in \mathfrak{S}_d$ is given by
\begin{equation}
	\begin{split}
		(\sigma,\tau)S_{p,q}^{(i,j)}=&(\sigma,\tau)S_{p,q-1}^{(i,j)}-\bigg(\langle \rho_{\sigma(p)}^{E_i}|\rho_{\sigma(p)}^{E_i}\rangle + \langle \rho_{\tau(q)}^{F_j}|\rho_{\tau(q)}^{F_j}\rangle)\bigg)\\
		&+\bigg(\langle \rho_{\sigma(p)}^{E_i}| \rho_{\sigma(p)}^{F_j}\rangle +\langle \rho_{\tau(q)}^{E_i}| \rho_{\tau(q)}^{F_j}\rangle\bigg)^2,\\
		(\sigma,\tau)	S_{p+1,1}^{(i,j)}=	&(\sigma,\tau)S_{p,p-1}^{(i,j)}-
		\bigg(\langle \rho_{\sigma(p+1)}^{E_i}|\rho_{\sigma(p+1)}^{E_i}\rangle + \langle \rho_{\tau(1)}^{F_j}|\rho_{\tau(1)}^{F_j}\rangle\bigg)\\
		&+\bigg(\langle \rho_{\sigma(p+1)}^{E_i}|\rho_{\sigma(p+1)}^{F_j}\rangle +\langle \rho_{\tau(1)}^{E_i}| \rho_{\tau(1)}^{F_j}\rangle\bigg)^2,
	\end{split}
\end{equation}
where $ 1\leq q< p \leq d$. It is evident that $S_{1,0}^{(i,j)}=I(\rho,E_i)I(\rho,F_j)$ is stable while $\mathfrak{S}_d\times \mathfrak{S}_d$ is present. Subsequently,
\begin{equation}\label{eq:Sij2}
	\begin{split}
	S_{1,0}^{(i,j)}&=(\sigma,\tau)S_{1,0}^{(i,j)}\geq 	(\sigma,\tau) S^{(i,j)}_{2,1}\geq 	(\sigma,\tau)S_{3,1}^{(i,j)}\geq 	(\sigma,\tau) S_{3,2}^{(i,j)}\\
		&\geq 	(\sigma,\tau)S_{4,1}^{(i,j)}\geq 	(\sigma,\tau)S_{4,2}^{(i,j)}\geq 	(\sigma,\tau)S_{4,3}^{(i,j)}\geq  \cdots \geq 	(\sigma,\tau) S^{(i,j)}_{d,d-1}.	
	\end{split}
\end{equation}
By adding $i$ from 1 to $n_1$ and $j$ from 1 to $n_2$, the following is obtained.
\begin{equation}
	\begin{split}
		I(\rho,\mathcal{N}_1)I(\rho,\mathcal{N}_2)&=(\sigma,\tau) S_{1,0} \geq (\sigma,\tau)S_{2,1}=(\sigma,\tau)I_2\\
		&\geq (\sigma,\tau)S_{3,1}\geq  (\sigma,\tau)S_{3,2}=(\sigma,\tau)I_3 \\ &\geq (\sigma,\tau)S_{4,1}\geq (\sigma,\tau)S_{4,2}\geq (\sigma,\tau)S_{4,3}=(\sigma,\tau)I_4\\
		&\geq \cdots\geq  (\sigma,\tau) S_{p,q}\geq \cdots\geq  (\sigma,\tau) S_{d,d-1}=(\sigma,\tau) I_d ,
	\end{split}
\end{equation}
where $(\sigma,\tau) S_{p,q}=\sum_{i=1}^{n_1}\sum_{j=1}^{n_2}(\sigma,\tau) S_{p,q}^{(i,j)}$.
Therefore, we get the following stronger result by optimizing over the symmetric group  $\mathfrak{S}_d$.
\begin{cor}
Let $\rho$ be a quantum state on a $d$-dimensional Hilbert space $H$. We have the following uncertainty sequence for two quantum channels $\mathcal{N}_1$ and $\mathcal{N}_2$ with the Kraus decompositions, $\mathcal{N}_1(\rho )=\sum_{i=1}^{n_1}E_i\rho E_i^{\dagger}$ and $\mathcal{N}_2(\rho )=\sum_{j=1}^{n_2}F_j \rho F_j^{\dagger}$, respectively.
	\begin{equation}\label{eq:th5}
		\begin{split}
			I(\rho,\mathcal{N}_1)I(\rho,\mathcal{N}_2)&\geq (1-t)S_{1,0}+t\max_{\sigma,\tau\in \mathfrak{S}_d} \{(\sigma,\tau) S_{p,q}\}\\
			&\geq \frac{1}{4}\sum_{i=1}^{n_1}\sum_{j=1}^{n_2}  \big([\sqrt{\rho},E_i], [\sqrt{\rho},F_j]\big)^2 ;\\
				I(\rho,\mathcal{N}_1)+I(\rho,\mathcal{N}_2)&\geq (1-t)S+t\max_{\sigma,\tau\in \mathfrak{S}_d}\{2\sqrt{(\sigma,\tau) S_{p,q}}\}\\
				&\geq \bigg( \sum_{i=1}^{n_1}\sum_{j=1}^{n_2} \big([\sqrt{\rho},E_i], [\sqrt{\rho},F_j]\big)^2\bigg)^{\frac{1}{2}},\\
		\end{split}
		\end{equation}
	where $0\leq t \leq 1$,   $1\leq q<p\leq d$, and  $  S=I(\rho,\mathcal{N}_1)+I(\rho,\mathcal{N}_2)$.
\end{cor}
\emph{Example 2.}
The quantum state  $\rho(\theta)$ is given by (\ref{rho}), the quantum channels $\mathcal{N}_1$ and $\mathcal{N}_2$ with the Kraus representations $\{E_i\}$ and $\{F_j\}$ are given by (\ref{K}). The improved product-form quantum channel uncertainty relations by skew information is given on the left of figure 4 and the sum-form is given on the right.
\begin{figure}
	\includegraphics[width=8cm,height=5cm]{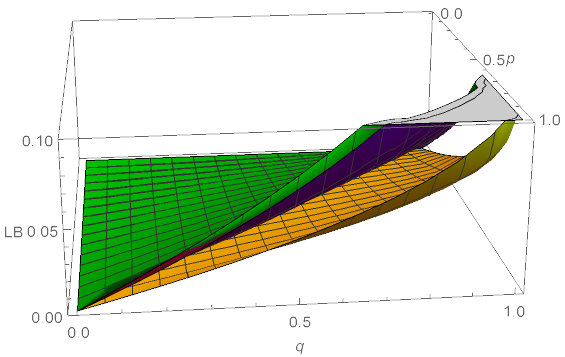}\quad 
	\includegraphics[width=8cm,height=5cm]{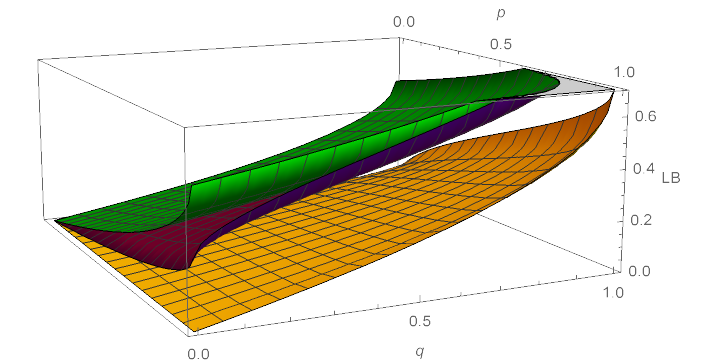}
	\caption{(color online) \textbf{The improved lower bounds of $I(\rho, \mathcal{N}_1)I(\rho,\mathcal{N}_2)$ (the left figure) and
$I(\rho, \mathcal{N}_1)+I(\rho,\mathcal{N}_2)$ (the right figure) for the state $\rho$ with $\theta= 1$.} Here $p$ and $q$ are the parameters of the Kraus representations for $\mathcal{N}_1$ and $\mathcal{N}_2$ respectively. The green surfaces (the top surfaces) in left and right are the product $I(\rho, \mathcal{N}_1)I(\rho,\mathcal{N}_2)$ in (\ref{eq:MP}) and the sum $I(\rho, \mathcal{N}_1)+I(\rho,\mathcal{N}_2)$ in (\ref{eq:MS}) respectively,  the purple surfaces (the 2nd surfaces) are our corrsponding improved lower bounds for the produt and sum form in (\ref{eq:th5}), and the left and right yellow surfaces (the bottom surfaces) are the lower bounds LBZ and $2\sqrt{\hbox{LBZ}}$ in \cite{ZZ} respectively.}
	\end{figure}
\bigskip

In general, there are many Kraus decompositions for quantum channels. We now show that our uncertainty relation sequences of the quantum channels are independent of the Kraus representations.
\begin{proposition}
	Let  $\rho$ be any quantum state in a Hilbert space $H$. Set the quantum channels $\mathcal{N}_1$ and $\mathcal{N}_2$ with the Kraus decompositions $\mathcal{N}_1(\rho )=\sum_{i=1}^{n_1}E_i\rho E_i^{\dagger}$ and $\mathcal{N}_2(\rho )=\sum_{j=1}^{n_2}F_j \rho F_j^{\dagger}$, respectively. Then our uncertainty relations in theorems and corollaries are independent of the Kraus representations of the quantum channels.
\end{proposition}
\proofname:
It is known that two operator-sum representations
$	\sum_{s}K_s\rho K_s^{\dagger}$ and $\sum_{t}K^{'}_t\rho (K^{'}_t)^{\dagger} $
describe the same channel if and only if there exists an unitary matrix $U=[U_{st}]$ such that $K_t^{'}=\sum_s U_{ts}K_s$ \cite{WG}. Suppose there are unitary matrices $U$ and $V$ such that $E_i^{'}=\sum_{t=1}^{n_1} U_{it}E_t$ and $F_j^{'}=\sum_{s=1}^{n_2} V_{js}F_s$, then $\mathcal{N}_1(\rho)=\sum_{t=1}^{n_1}E_t\rho E_t^{\dagger}=\sum_{t=1}^{n_1}E_t^{'}\rho (E^{'}_t)^{\dagger} $ and  $\mathcal{N}_2(\rho)=\sum_{s=1}^{n_2}F_s\rho F_s^{\dagger}=\sum_{s=1}^{n_2}F_s^{'}\rho (F^{'}_s)^{\dagger} $. We show the lower bound $I_m$ is independent of the Kraus representations for quantum channels.
Recall that  $I_m=\sum_{i=1}^{n_1}\sum_{j=1}^{n_2}  I_m^{(i,j)}$ and $I_m^{(i,j)}$ are given in (\ref{Imij}). By the same construction process as $I_m^{(i,j)}$, let $\rho^{E_i^{'}}:=[\sqrt{\rho}, E_i^{'}]$ and
$\rho^{F_j^{'}}:=[\sqrt{\rho}, F_j^{'}]$, and
\begin{equation}
\begin{split}
	|\rho^{E_i^{'}}_k\ra:=\rho^{E_i^{'}}|k\ra&=\sum_tU_{it}\rho^{E_t}|k\ra=\sum_tU_{it}|\rho^{E_t}_k\ra,\\
	|\rho^{F_j^{'}}_l\ra:=\rho^{F_j^{'}}|l\ra&=\sum_sV_{is}\rho^{F_s}|l\ra=\sum_sV_{is}|\rho^{F_s}_l\ra.
\end{split}
\end{equation}
Setting
\begin{equation}
	\begin{split}
\vec{E^i}:=&	(|\rho^{E_i^{'}}_1\ra,\cdots,|\rho^{E_i^{'}}_d\ra)=(\sum_tU_{it}|\rho^{E_t}_1\ra,\cdots,\sum_tU_{it}|\rho^{E_t}_d\ra),\quad
\\ \vec{F^j}:=&(|\rho^{F_j^{'}}_l\ra,\cdots,|\rho^{F_j^{'}}_d\ra)=(\sum_sV_{is}|\rho^{F_s}_1\ra,\cdots,\sum_sV_{is}|\rho^{F_s}_d\ra).
	\end{split}
\end{equation}
Then the product
\begin{equation}
	\begin{split}
	I(\rho,E_i^{'})I(\rho,F_j^{'})=&(|\vec{E}^i_m|^2+|\vec{E}^i_{m,c}|^2)( |\vec{F}^j_m|^2+|\vec{F^j}_{m,c}|^2)\\
	\geq& |\lan \vec{E}^i_m,\vec{F}^j_m \ra |^2+|\vec{E^i}_{m,c}|^2|\vec{F^j}_{m,c}|^2+|\vec{E}^i_m|^2|\vec{F^j}_{m,c}|^2=:J_m^{(i,j)}.
\end{split}
\end{equation}
The vectors $\vec{E}^i_m$ and $\vec{F}^j_m$ are constructed using the same definition as $\vec{e}^{~i}_m$ and $\vec{f}^{~j}_m$, then
\begin{equation}
		\begin{split}
&\sum_{i=1}^{n_1}\sum_{j=1}^{n_2}|\lan \vec{E}^i_m,\vec{F}^j_m \ra |^2=\sum_{i=1}^{n_1}\sum_{j=1}^{n_2}\lan \vec{E}^i_m,\vec{F}^j_m \ra \lan \vec{F}^j_m,\vec{E}^i_m\ra \\
	=&\sum_{i=1}^{n_1}\sum_{j=1}^{n_2}\bigg(\sum_{t=1}^{n_1}U^{\dagger}_{it} \cdot \sum_{s=1}^{n_2} V_{js}\big(\sum_{k=1}^m\lan\rho^{E_t}_k|\rho^{F_s}_k\ra\big)\bigg)
	\bigg(\sum_{s^{'}=1}^{n_2}V^{\dagger}_{js^{'}} \cdot \sum_{t^{'}=1}^{n_1} U_{it^{'}}\big(\sum_{l=1}^m\lan\rho^{F_{s^{'}}}_{l}|\rho^{E_{t^{'}}}_{l}\ra\big)\bigg)\\
	=&\sum_{t=1}^{n_1}\sum_{s=1}^{n_2} \bigg(\sum_{i=1}^{n_1}\sum_{t^{'}=1}^{n_1}U^{\dagger}_{it}\bigg(\sum_{j=1}^{n_2}\sum_{s^{'}=1}^{n_2}V_{js}V^{\dagger}_{js^{'}}\bigg)U_{it^{'}}\bigg)\bigg(\sum_{k=1}^m\lan \rho^{E_t}_k|\rho^{F_s}_k\ra\bigg)\bigg(\sum_{l=1}^m\lan \rho^{F_{s^{'}}}_l|\rho^{E_{t^{'}}}_l\ra\bigg)\\
	=& \sum_{t=1}^{n_1}\sum_{s=1}^{n_2}|\sum_{k=1}^m\lan \rho^{E_t}_k|\rho^{F_s}_k\ra|^2
		=\sum_{i=1}^{n_1}\sum_{j=1}^{n_2}|\lan\vec{e}^{~i}_m,\vec{f}^{~j}_m\ra|^2,
	\end{split}
\end{equation}
 Similarly, we can show that \begin{equation}
 	\begin{split}
\sum_{i=1}^{n_1}\sum_{j=1}^{n_2}|\vec{E}^i_{m,c}|^2|\vec{F^j}_{m,c}|^2&=\sum_{i=1}^{n_1}\sum_{j=1}^{n_2}|\vec{e}^{~i}_{m,c}|^2|\vec{f}^{~j}_{m,c}|^2,\\
 \sum_{i=1}^{n_1}\sum_{j=1}^{n_2}|\vec{E}^i_m|^2|\vec{F^j}_{m,c}|^2&=\sum_{i=1}^{n_1}\sum_{j=1}^{n_2}|\vec{e}^{~i}_m|^2|\vec{f}^{~j}_{m,c}|^2.
 \end{split}\end{equation} Therefore, we have verified that
\begin{equation}
J_m:=\sum_{i=1}^{n_1}\sum_{j=1}^{n_2}	J_m^{(i,j)}=\sum_{i=1}^{n_1}\sum_{j=1}^{n_2}I_m^{(i,j)}=:I_m.
\end{equation}
Similarly the lower bounds $S_{p,q}$ in (\ref{eq:th3}) can be seen independent from the choice of the Kraus operators for quantum channels.
Therefore the uncertainty relations are independent of the quantum channel Kraus representations.

\section{\label{sec:leve4} Conclusion}
As we known, almost all quantum cryptographic protocols, including quantum key distribution and two-party quantum cryptography, have the uncertainty relation as a fundamental component of their security analysis \cite{CB}. The uncertainty relation is also known as an effective tool in quantum metrology \cite{GV}, entanglement witness \cite{BC}, EPR steering \cite{WS, SB}, and quantum random number creation \cite{ CZ, VM}. In this article, we have proposed enhanced product-form and sum-form uncertainty relation sequences for quantum channels based on skew information.  Our lower bounds are significantly better than some of the recently known results. In our method, the observables are measured with sampling measurement coordinates to improve the Cauchy-Schwarz inequality. Moreover, we have shown that the uncertainty relation sequences are independent of the Kraus representations for quantum channels. Since most existing uncertainty relations for quantum channels depend on the Kraus representations, they can be strengthened by optimizing across all feasible Kraus representations. From an alternative perspective, it would be beneficial to seek uncertainty relations for quantum channels that are invariant under the Kraus representations. We anticipate that these findings will be helpful for future research on the uncertainty relations for arbitrary quantum channels as well as quantum information in general.

\bigskip

\centerline{\bf Acknowledgments}
The research is supported in part by NSFC grants 12226321, 12226320, 12171303 and 11871325. It is also supported by theResearch Fund of JianghanUniversity (No. 2023JCYJ08), the Science and
Technology Commission Of Shanghai Municipality (No. 22DZ2229014), Guangdong Basic and Applied
Basic Research Foundation 2020A1515111007 and the Start-up Fund of Guangdong Polytechnic Normal
University no. 2021SDKYA178.
 \vskip 0.1in

\textbf{Data Availability Statement.} All data generated during the study are included in the article.

\textbf{Conflict of Interest Statement.} The authors declare that the research was conducted in the
absence of any commercial or financial relationships that could be construed as a
potential conflict of interest.
\bibliographystyle{amsalpha}

\begin{thebibliography}{ABCD}
		
\bibitem{1} Fuchs C.A. and Peres A.: Quantum-state disturbance versus information gain: Uncertainty relations for quantum information. Phys. Rev. A $\mathbf{53}$, 2038 (1996)

\bibitem{RB} Renes J.M. and Boileau J.C.: Conjectured strong complementary information tradeoff. Phys. Rev. Lett. $\mathbf{103}$, 020402 (2009)


\bibitem{3} Bowen W.P., Schnabel R., Lam P.K. and Ralph T.C.: Experimental investigation of criteria for continuous variable entanglement.
Phys. Rev. Lett. $\mathbf{90}$, 043601 (2003)

\bibitem{4} G\"uhne O.: Characterizing entanglement via uncertainty relations. Phys. Rev. Lett. $\mathbf{92}$, 117903 (2004)

\bibitem{5} Howell J.C., Bennink R.S., Bentley S.J. and Boyd R.W.: Realization of the Einstein-Podolsky-Rosen Paradox Using Momentum- and Position-Entangled Photons from Spontaneous Parametric Down Conversion. Phys. Rev. Lett. $\mathbf{92}$, 210403 (2004)

\bibitem{6} Pires D.P., Cianciaruso M., C\'eleri L.C., Adesso G. and Soares-Pinto D.O.: Generalized geometric quantum speed limits. Phys. Rev. X $\mathbf{6}$, 021031(2016)

\bibitem{7}  Cand\'es E.J., Romberg J. and Tao T.:Robust uncertainty principles: Exact signal reconstruction from highly incomplete frequency information. IEEE. Trans. Inf. Theory $\mathbf{52}$, 489(2006)

\bibitem{15} Heisenberg W.: \"Uber den anschaulichen Inhalt der quantentheoretischen Kinematik und Mechanik. Z. Phy. $\mathbf{43}$, 172 (1927)

\bibitem{16} Kennard E.H.: Zur quantenmechanik einfacher bewegungstypen. Z. Phys. $\mathbf{44}$, 4 (1927)

\bibitem{17} Weyl H.: Gruppentheorie und Quantenmechanik, Hirzel, Leipzig (1928)

\bibitem{18} Robertson H.P.: The uncertainty principle. Phys. Rev. $\mathbf{34}$, 163 (1929)

\bibitem{GH} Grudka, A., Horodecki, M., Horodecki, P., Horodecki, R., Klobus, W., Pankowski, L: Conjectured strong complementary-correlations tradeoff. Phys. Rev. A. $\mathbf{88}$, 032106 (2013)

\bibitem{DD} Deutsch, D.: Uncertainty in quantum measurements. Phys. Rev. Lett. $\mathbf{50}$, 631 (1983)

\bibitem{MHU} Maassen, H., Uffink, J.B.M.: Generalized entropic uncertainty relations. Phys. Rev. Lett. $\mathbf{60}$, 1103 (1988)

\bibitem{WY} Wu, S., Yu, S., M\o lmer, K.: Entropic uncertainty relation for mutually unbiased bases. Phys. Rev. A. $\mathbf{79}$, 022104 (2009)

\bibitem{GG} Gour, G., Grudka, A., Horodecki, M., Klobus, W., Lodyga, J., Narasimhachar, V.: Conditional uncertainty principle. Phys. Rev. A. $\mathbf{97}$, 042130 (2018)

\bibitem{BG} Busch, P., Grabowski, M., Lahti, P.: Operational quantum physics. Springer, Berlin (1995)

\bibitem{NC} Nielsen, M.A., Chuang, I.L.: Quantum computation and quantum information, 10th edn. Cambride University Press, Cambridge (2010)

\bibitem{XJ} Xu, J.: Coherence of quantum channels. Phys. Rev. A. $\mathbf{100}$, 052311 (2019)

\bibitem{MK} Mani, A., Karimipour, V.: Cohering and decohering power of quantum channels. Phys. Rev. A. $\mathbf{92}$, 032331 (2015)

\bibitem{TRA} Takahashi, M., Rana, S., Streltsov, A.: Creating and destroying coherence with quantum channels.Phys. Rev. A. $\mathbf{105}$, 060401 (2022)

\bibitem{FSL} Fu, S., Sun, Y., Luo, S.: Skew information-based uncertainty relations for quantum channels. Quantum Inf. Process. $\mathbf{18}$, 258 (2019)

\bibitem{LS18} Luo, S., Sun, Y.: Coherence and complementarity in state-channel interaction. Phys. Rev.A. $\mathbf{98}$, 012113 (2018)

\bibitem{ZWF} Zhang, Q.H., Wu, J.F., Fei, S.M.: A note on uncertainty relations of arbitrary $N$ quantum channels. Laser  Phys. Lett. $\mathbf{18}$, 095204 (2021)

\bibitem{LS05} Luo, S.: Quantum versus classical uncertainty. Thero. Math. Phys. $\mathbf{143}$, 681 (2005)

\bibitem{ZG} Zhang, L.M., Gao, T. and Yan, F.L.: Tighter uncertainty relations based on Wigner-Yanase skew information for observables and channels. Phys. Lett. A. $\mathbf{387}$, 127029 (2021)
\bibitem{LG} Li H., Gao, T. and Yan, F.L.: Tighter sum uncertainty relations via metric-adjusted skew
information. Phys. Scr. $\mathbf{98}$, 015024 (2023)

\bibitem{YC} Yu, C.: Quantum coherence via skew information and its polygamy. Phys. Rev. A. $\mathbf{95}$, 042337 (2017)

\bibitem{LS03} Luo, S.: Wigner-Yanase skew information and uncertainty relations. Phys. Rev. Lett. $\mathbf{91}$, 180403 (2003)

\bibitem{ZZ}  Zhou N., Zhao M., Wan Z. G,  Li T.: The uncertainty relation for quantum channels based on skew information. Quantum Infor. Process $\mathbf{22}$, 6 (2023)

\bibitem{SY} Luo, S., Sun, Y.: Coherence and complementarity in state-channel interaction. Phys. Rev. A $\mathbf{98}$, 012113
(2018)

\bibitem{HJ23} Hu X., Jing N.: Uncertainty relations for metric-adjusted skew information and Cauchy-Schwarz inequality. Laser Phys. Lett. $\mathbf{20}$, 085202 (2023)

\bibitem{CB} Coles, P.J., Berta, M., Tomamichel, M., Wehner, S.: Entropic uncertainty relations and their applications. Rev. Mod. Phys. $\mathbf{89}$, 015002 (2017)

\bibitem{GV} Giovannetti, V., Lloyd, S., Maccone, L.: Advances in quantum metrology. Nat. Photon. $\mathbf{5}$, 222 (2011)

\bibitem{BC} Berta, M., Coles, P.J., Wehner, S.: Entanglement-assisted guessing of complementary measurement outcomes. Phys. Rev. A $\mathbf{90}$, 062127 (2014)

\bibitem{WS} Walborn, S.P., Salles, A., Gomes, R.M., Toscano, F., Ribeiro, P.H.S.: Revealing hidden Einstein-Podolsky-Rosen nonlocality. Phys. Rev. Lett. $\mathbf{106}$, 130402 (2011)

\bibitem{SB} Schneeloch, J., Broadbent, C.J., Walborn, S.P., Cavalcanti, E.G., Howell, J.C.: Einstein-Podolsky-Rosen steering inequalities from entropic uncertainty relations. Phys. Rev. A $\mathbf{87}$, 062103 (2013)

\bibitem{CZ} Cao, Z., Zhou, H., Yuan, X., Ma, X.: Source-independent quantum random number generation. Phys. Rev. X $\mathbf{6}$, 011020 (2016)

\bibitem{VM} Vallone, G., Marangon, D.G., Tomasin, M., Villoresi, P.: Quantum randomness certified by the uncertainty principle. Phys. Rev. A $\mathbf{90}$, 052327 (2014)

\bibitem{WG} Ritter W.G.: Quantum channels and representation theory. J. Math. Phys.  $\mathbf{46}$, 082103 (2005)

\end{thebibliography}

\end{document}